\documentclass[aps,twocolumn]{revtex4}
\usepackage{amssymb}
\usepackage{graphicx}
\usepackage{dcolumn}
\usepackage{bm}

\begin{document}

\title{Strategies for Real-Time Position Control of a Single Atom in Cavity
QED}
\author{T. W. Lynn, K. Birnbaum, H. J. Kimble}
\affiliation{Norman Bridge Laboratory of Physics 12-33, California Institute of
Technology, Pasadena, CA 91125, USA}
\date{July 6, 2005}

\begin{abstract}
Recent realizations of single-atom trapping and tracking in cavity QED open
the door for feedback schemes which actively stabilize the motion of a
single atom in real time. We present feedback algorithms for cooling the
radial component of motion for a single atom trapped by strong coupling to
single-photon fields in an optical cavity. Performance of various algorithms
is studied through simulations of single-atom trajectories, with full
dynamical and measurement noise included. Closed loop feedback algorithms
compare favorably to open-loop ``switching" analogs, demonstrating the
importance of applying actual position information in real time. The high
optical information rate in current experiments enables real-time tracking
that approaches the standard quantum limit for broadband position
measurements, suggesting that realistic active feedback schemes may reach a
regime where measurement backaction appreciably alters the motional dynamics.
\end{abstract}

\maketitle


\section{Introduction}

Recent experiments \cite{acm,pinkse,fort,mckeever,mcknature} have
demonstrated the ability to trap \cite%
{acm,fort,mckeever,mcknature,rempe-nature} and localize
\cite{pinkse} a single atom in a high-finesse optical cavity by
way of optical forces. Moreover, by detecting the light
transmitted by the cavity, the atom's motion within the cavity
mode can be monitored in real time with high signal to noise
throughout its trapped lifetime \cite{acdpra}.These achievements
in trapping and localization open up exciting possibilities for
quantum logic and quantum state preparation in the context of cavity QED\cite%
{cirac98,pcz,cirac97,vanenk98,briegelrep,briegelnet}. Beyond the
realization of trapping, the high signal-to-noise for continuous,
real-time position measurement is itself one of the most notable
features of these strongly coupled cavity QED systems. Such
detailed real-time position information immediately suggests the
idea of active feedback to dynamically cool the motion of a single
trapped atom. By investigating this system and the feedback
schemes available in it, basic questions of quantum state
estimation and optimal control can be explored for continuous
measurement of a dynamical variable -- in this case the position
of a single atom.

Crucial to the realization of trapping and sensing in cavity QED is \textit{%
strong coupling}, a condition in which the coherent coupling between atom
and cavity field dominates dissipative rates in the system. For a two-state
atom optimally coupled to the cavity mode, the dipole-field coupling is
given by the Jaynes-Cummings interaction Hamiltonian \cite{jc}
\begin{equation}
H_{int}=\hbar g(\sigma _{+}a+\sigma _{-}a^{\dag }),  \label{Hint}
\end{equation}%
where $\sigma _{\pm }$ are dipole raising and lowering operators, $%
(a,a^{\dag })$ are field annihilation and creation operators for the cavity
mode, and $g$ is one half of the single-photon Rabi frequency. This
interaction gives rise to the well-known Jaynes-Cummings ladder of
eigenstates for the coupled atom-cavity system, and correspondingly to the
vacuum Rabi splitting for the system's resonant frequencies \cite{eberly}.
Dissipation, on the other hand, is characterized by the cavity decay rate $%
\kappa $ and the atomic spontaneous emission rate $\gamma $. Strong coupling
occurs for $g\gg (\kappa ,\gamma )$. We can define a further condition of
strong coupling for the \textit{external} atomic degrees of freedom; this
occurs when the coherent coupling also dominates the atomic kinetic energy,
as first achieved in Ref. \cite{HoodPRL}.

Under these conditions, interaction with a single-photon cavity
field exerts a strong mechanical effect on a single atom, allowing
trapping of the atom when the system is driven to strong-field
seeking states in the Jaynes-Cummings ladder. Furthermore, strong
coupling assures that the intracavity light field and thus the
cavity output field (transmitted light) are influenced by an atom;
thus, as an atom moves between more and less strongly coupled
positions within the cavity, the transmitted light provides a
real-time measurement of atomic position. This sensing enables the
one-time triggering employed in Refs.
\cite{acm,mckeever,pinkse,fort} to switch on a trapping potential
when an atom is present near the center of the cavity. The ongoing
stream of position information should, however, be useful for
continued active feedback based on the atomic position. Rather
than simply triggering a potential to turn on, it should be
possible to modulate the potential depth to dynamically cool the
atomic motion, in a method analogous to the principles of
stochastic cooling but for a single atom. Initial steps in this
direction are represented in the work of Ref. \cite{rempefb}.
Quantum feedback for other atomic state variables is also an
active area of research, including the recent experimental
demonstration of quantum feedback to control the ensemble spin of
a collection of cold atoms \cite{hmspinexpt}.

In this paper, we address the question of how to implement atomic
position feedback in experimentally realistic situations, where
several constraints apply: (1) the system is inherently nonlinear
and largely nonanalytic, with relationships between many
quantities of interest determined by steady-state solutions to the
master equation for the atom-cavity system, (2) dynamical noise is
significant and changes in tandem with the driving field and
trapping potential, and (3) measurement noise, arising largely
from the fundamental quantum noise (shot noise) of detection,
imposes necessary delays in the estimation of dynamical variables
and the implementation of feedback. Broadband measurement near the
standard quantum limit (see Ref. \cite{sql}) has been demonstrated in this system
through measurement of cavity transmission amplitude \cite{acm}
and by simultaneous measurement of transmitted amplitude and phase
\cite{fullmeas}. Use of these measurements in feedback control of
some aspect of the atomic motion should bring us closer to regimes
where measurement backaction has a significant effect, so that
different detection methods may exhibit different control limits
based on these effects as well as on more conventional
signal-to-noise considerations \cite{andreis,gambetta}.

In Section II we review the experimental system on which our work is based.
The feedback strategy we consider is presented and motivated in Section III.
Section IV presents feedback results; these are considered both in the realistic
experimental context and in several idealized systems in order to illustrate the
algorithm's effect on the individual components of atomic motion.  In Section V
we again take up the topic of experimentally measurable feedback results, and in
Section VI we discuss limits and possible extensions of the algorithms discussed
in the paper.

\section{Experimental Status and Sensitivity for Atomic Position Measurement}

In our feedback calculations, we have considered the situation of Ref. \cite%
{acm}, where a single atom is trapped via its interaction with a
near-resonant cavity light field at the level of a single photon. In that
experiment, the small saturation photon number and critical atom number \cite%
{hjkscripta} -- the smallest in experiments to date -- facilitated
not only the trapping mechanism but, more importantly for these
considerations, the high signal-to-noise observation of atomic
motion within the cavity field.

\begin{figure}[tbph]
\includegraphics[width=8.6cm]{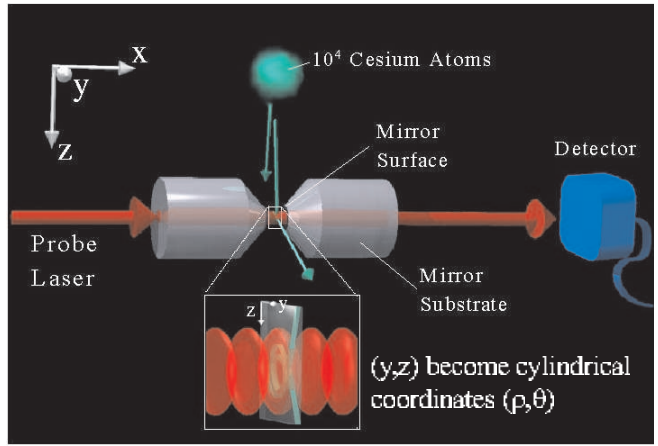}
\caption{Cesium atoms are dropped from a MOT and a small fraction of these
fall one by one through the mode of an optical cavity. The cavity mode is a
cylindrically symmetric Gaussian in the transverse $(\protect\rho ,\protect%
\theta )$ plane and has a standing-wave structure in the axial (x)
direction. Transmission of a probe beam through the cavity is monitored to
sense an atom's motion through the cavity mode; a detected rise in
transmission triggers a switch to higher probe intensity, significantly
populating strong-field-seeking states of the atom-cavity system and thus
trapping the atom. The figure shows one falling atom which misses the cavity
mode altogether and a second atom which enters the cavity, is trapped in the
mode, and eventually escapes.}
\label{fig:expt}
\end{figure}

Figure \ref{fig:expt} shows a schematic diagram of the experimental
procedure, in which Cesium atoms (atomic resonance frequency $\omega_{a}$)
are collected in a magneto-optical trap (MOT) directly above the cavity
mirrors, cooled to temperatures of $\sim 10 \mu$K, and dropped through the
cavity. The geometry of the mirror substrates cuts off most of the atomic
flux so that one atom at a time transits the cavity mode of length $l$ and
Gaussian waist $w_{0}$. The cavity resonance $\omega_{c}$ is tuned near but
slightly below the atomic resonance frequency so that $\omega_{c}-%
\omega_{a} < 0$. The cavity is continuously driven by a probe
laser at frequency $\omega_{p}$, and the transmission of this beam
through the cavity is monitored via balanced heterodyne detection.
For a probe red-detuned from both atom and cavity ($\omega_{p}$
near the lower vacuum Rabi sideband\cite{eberly}), transmission is
low for the empty cavity and is highest when an atom is in the
regions of strongest coupling. Thus the monitored photocurrent
carries real-time information about the atomic position, with high
signal-to-noise even for probe strengths corresponding to $<1$
intracavity photons. Saturation of the atom-cavity
response\cite{HoodPRL} sets in for larger field strength, so that
the most sensitive tracking is realized at $\sim 1$ intracavity
photon.

The atom-cavity evolution in this system is described by a master equation
(see, e.g., \cite{api,masterequation,tan1999a}) for the joint atom-cavity
density operator $\rho$. We consider a driving (and probing) field $\epsilon$
of frequency $\omega_{p}$, a cavity resonant at $\omega_{c}=\omega_{p}+%
\Delta_{cp}$, and an atomic resonance frequency $\omega_{a}=\omega_{p}+%
\Delta_{ap}$. In the electric dipole and rotating-wave
approximations, and in the interaction picture with respect to the
probe frequency, the master equation can be written
\begin{eqnarray}
\dot{\hat{\rho}}=-\frac{i}{\hbar}[\hat{H}_{0},\hat{\rho}]+ \gamma(2\hat{\sigma}%
\hat{\rho}\hat{\sigma}^{\dag}-\hat{\sigma}^{\dag}\hat{\sigma}\hat{\rho}-\hat{\rho}\hat{\sigma}%
^{\dag}\hat{\sigma})  \nonumber \\
+\kappa(2\hat{a}\hat{\rho}\hat{a}^{\dag}-\hat{a}^{\dag}\hat{a}\hat{\rho}-\hat{\rho}\hat{a}%
^{\dag}\hat{a}),  \label{eq:meq} \\
\hat{H}_{0}=\hbar\Delta_{cp}\hat{a}^{\dag}\hat{a}+\hbar\Delta_{ap}\hat{\sigma%
}^{\dag}\hat{\sigma} + \hbar g(\vec{r})[\hat{a}\hat{\sigma}^{\dag}+\hat{a}%
^{\dag}\hat{\sigma}]  \nonumber \\
+ \hbar\epsilon(\hat{a}+\hat{a}^{\dag}).  \label{eq:meq2}
\end{eqnarray}
Here $g(\vec{r})$ is the coupling strength which takes into account the
atomic position $\vec{r}$ within the cavity mode. For a Fabry-Perot cavity
supporting a standing wave mode with Gaussian transverse profile, $g(\vec{r}%
) = g_{0}cos(2\pi x/\lambda)exp[-(y^2+z^2)/w_{0}^2]$. The
cylindrical symmetry of the field suggests the use of cylindrical
coordinates $(\rho,\theta,x)$, where $\rho=\sqrt{y^2+z^2}$ and
$\theta=tan^{-1}(-z/y)$ (see Figure \ref{fig:expt}). Thus we write
$g(\vec{r})=g_{0}cos(2\pi x/\lambda)exp[-\rho^2/w_{0}^2]$.  In the
fully quantum treatment, the atomic position $\vec{r}$ is itself
an operator; in experiments to date a quasi-classical treatment
suffices, so the atom may be considered a wavepacket with
$\vec{r}$ a classical center-of-mass position variable. Similar
feedback schemes for an atom-cavity system have also been explored
theoretically in the case of an atom which has already been cooled
radially and must now be treated in a fully quantized manner for
cooling of the remaining axial motion\cite{steckfb}.

Following the experimental situation of Ref.\cite{acm}, we
consider an atom-cavity system in which
$(g_{0},\protect\kappa,\protect\gamma)/2\protect\pi =
(110,14.2,2.6)$ MHz.  The simulation results below refer to
varying cavity field strength but with detunings fixed at
$(\protect\omega_{c}-\protect\omega_{a})/2\protect\pi = -47$ MHz,
$(\protect \omega_{p}-\protect\omega_{a})/2\protect\pi = -125$
MHz.

\begin{figure}[tbph]
\includegraphics[width=8.6cm]{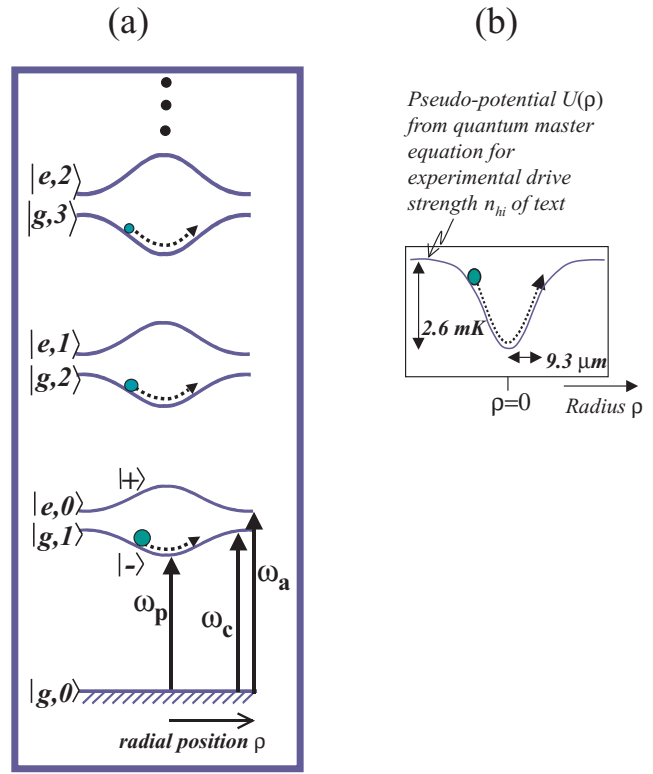}
\caption{(a)First few levels of the ladder of energy eigenstates
for the atom-cavity system. Uncoupled and maximally coupled
eigenvalues are illustrated, along with their smooth dependence on
the atomic position (and hence the atom-cavity coupling). Strength
of driving light determines the system's population distribution
across the first few levels of the ladder, setting the shape and
depth of the effective potential. (b) Effective potential for the
atom, cavity, and probe detuning parameters of the simulations,
and drive strength $n_{hi}$ corresponding to 0.3 photons in the
empty cavity on resonance. } \label{fig:ladderpot}
\end{figure}

Figure \ref{fig:ladderpot}(a) shows the first few levels of the
Jaynes-Cummings ladder of energy eigenvalues, obtained by
diagonalizing the interaction Hamiltonian of Eq. \ref{eq:meq2}.
The smooth evolution from uncoupled to fully coupled eigenstates
reflects the dependence of coupling $g(\vec{r})$
on the atomic position $\vec{r}$, and specifically on the atom's distance $%
\rho$ from the cavity axis. In the presence of dissipation and
driving, the distribution of populations across the first few
levels of this ladder is determined by numerical steady-state
solution of the master equation at each position. Figure
\ref{fig:ladderpot}(b) shows the effective potential for the
atom-cavity properties considered in this work, with driving level
in this example fixed at $n_{hi}$=0.3 photons in the empty cavity.
The atom-trapping scheme of \cite{acm} is based on tracking the
atomic position and altering the driving field strength to place
the system in the attractive potential of Figure
\ref{fig:ladderpot}(b) when the atom is close to the cavity axis
($\rho\approx 0$).

\begin{figure}[tbph]
\centerline{(a)}\centerline{%
\includegraphics[width=8.6cm]{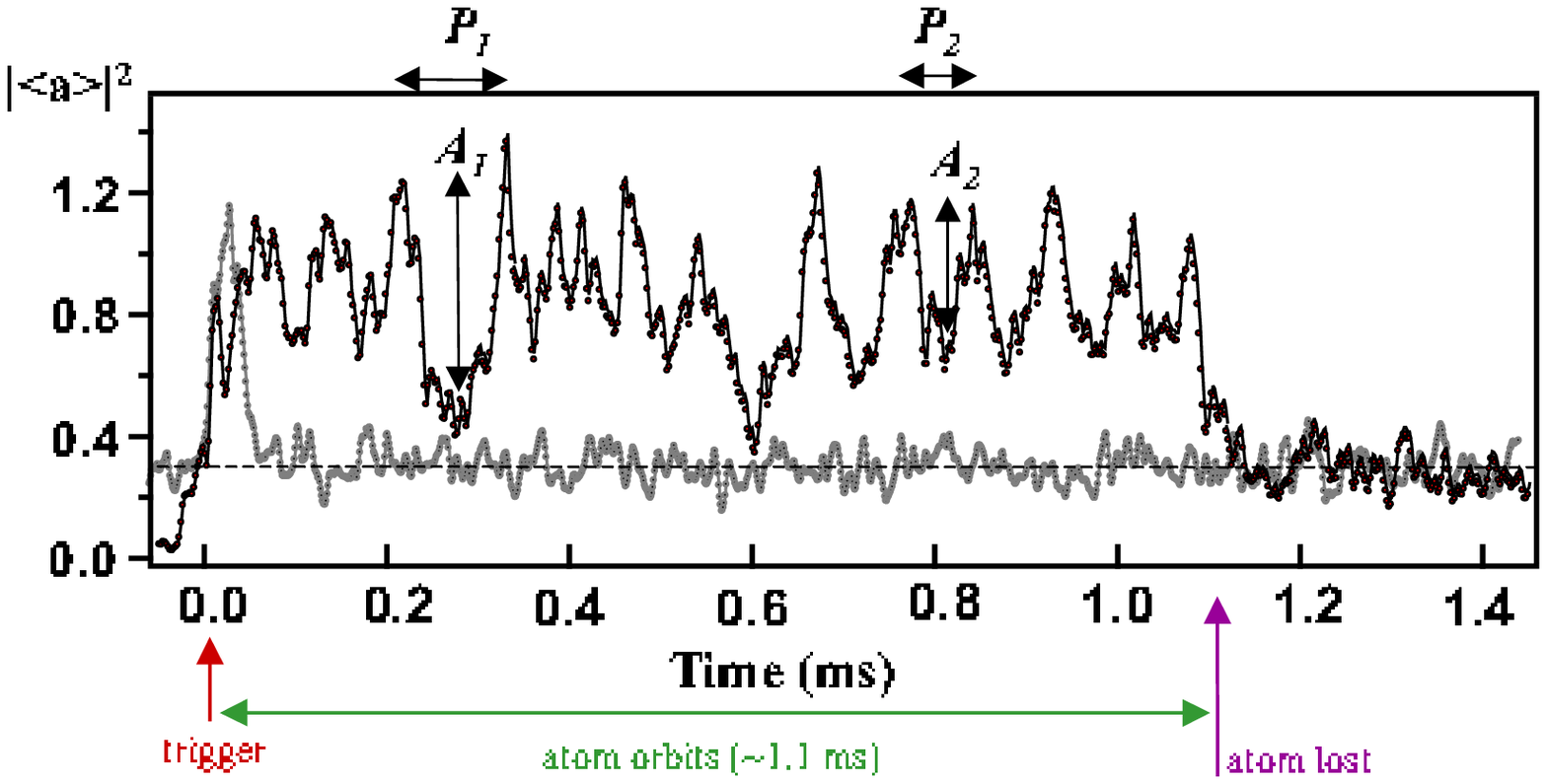}} \vspace{0.2in} %
\centerline{(b)}\centerline{\includegraphics[width=8.6cm]{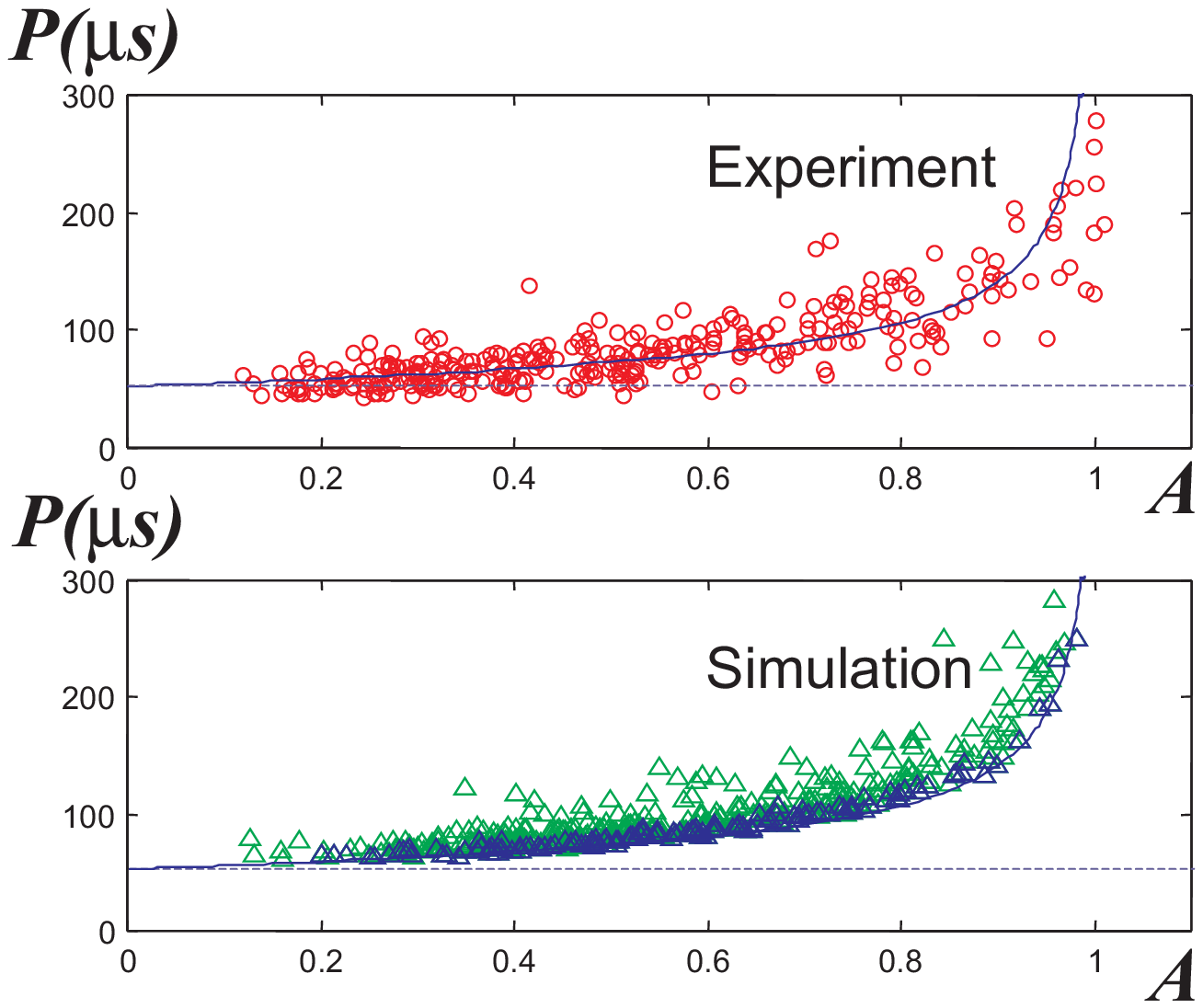}}
\caption{(a) Cavity transmission record for an atom trapped with
0.3 photons in the empty cavity. Motion of the trapped atom within
the cavity mode can be tracked via the oscillations in cavity
transmission. Each oscillation has amplitude A and period P as
indicated on the trace. (b) Oscillation period P vs. amplitude A
of transmission oscillations as in (a). The solid curve is
calculated for motion in the known anharmonic potential of Figure
\protect \ref{fig:ladderpot}(b), with no free parameters in the
fit.} \label{fig:trappedtransit}
\end{figure}

Figure \ref{fig:trappedtransit}(a) shows a sample experimental
trace of transmission vs. time for a single atom trapped as in
Ref. \cite{acm}. Immediately notable in the transmission record
are the large, regular oscillations in the heterodyne current,
which can be associated with an atom repeatedly approaching and
receding from the regions of strongest atom-cavity coupling. The
cavity mode structure produces a trapping potential with width
$\sim\lambda/2=426$nm in the axial direction and $\sim
w_{0}=14\mu$m in the radial direction, giving axial oscillations
at $\sim 1-2$MHz while radial oscillations occur on the much
slower timescale of $\sim 10-20$kHz. By using a detection
bandwidth of 100kHz, we obtain a transmission record that averages
over the effects of axial motion.

A series of quantitative comparisons demonstrates that the remaining
transmission signal faithfully reflects radial motion with very little
contamination from the axial averaging\cite{acdpra}. Simulations indicate
that, for the parameter regime employed in Ref.\cite{acm}, a trapped atom is
typically confined within $\sim 50nm$ of a single standing-wave antinode
until it heats very quickly and escapes the trap altogether. Events
involving large-amplitude oscillations or `skipping' across wells and
retrapping in another antinode are very rare occurrences in this regime;
since axial motion typically has such small amplitudes, we expect the
average over it to have a negligible effect on measured cavity transmission.
Comparison of observed maximum transmission levels with the expected
theoretical maximum confirms this notion, giving an estimate of $\lesssim
75nm$ for typical axial excursions.

As further confirmation that transmission accurately reflects radial motion,
we consider the observed period of transmission oscillations as a function
of their amplitude. From our knowledge of the anharmonic (approximately
Gaussian) trapping potential in the radial direction, it is straightforward
to calculate the expected relationship between amplitude and period for
these oscillations. As evident from Figure \ref{fig:trappedtransit}(b),
actual data closely follow the theoretical curve, indicating that the
transmission record can indeed be interpreted as a record of radial motion ($%
T(t)\rightarrow \rho (t)$). Though the atom-cavity coupling is
cylindrically symmetric and thus provides no explicit information
about $\theta (t)$, knowledge of $\rho (t)$ and the trapping
potential allow us to reconstruct an estimate of an atom's angular
momentum $L(t)$ and thus of a two-dimensional trajectory in the
$(\rho ,\theta )$ plane. Such a method can be applied with success
in a parameter regime where the atomic motion is largely
conservative and the angular momentum varies slowly on the
timescale of a single radial orbit\cite{acm,acdpra}.

Three basic ambiguities will be clear from this algorithm for
trajectory reconstruction: 1) the sign of the angular momentum is
unknown, so the trajectory has arbitrary handedness. 2) the
initial angle $\theta_{0}$ is arbitrary, so the resulting
trajectory can be rotated freely as a unit. 3) The trajectory is
constructed in two dimensions, with the axial motion confined
within a single antinode, but no information is available about
\textit{which} antinode the atom occupies during the trajectory.
These ambiguities, while noted here for clarity, arise in aspects
of the motion not used in the feedback scheme treated below.

Two-dimensional trajectory reconstructions dramatically illustrate
the cavity-enhanced sensing power for atomic motion. However, the
initial goal of our feedback algorithms will be to control
$\rho(t)$; for this purpose it is sensible to ignore $\theta(t)$
and apply all available signal to noise to the task of estimating
$\rho(t)$ and $\dot{\rho}(t)$ in real time. The goal of such a
program is then to use this information to drive $\rho(t)$ to a
constant value, or in other words to circularize an orbit in the
$(\rho,\theta)$ plane while not necessarily driving it to the
cavity axis ($\rho=0$). The latter task, which requires an
explicit method of breaking cylindrical symmetry for position
sensing and for the effective potential, can be considered as a
later extension.

\section{The Atom and Cavity as a Control System: Basic Feedback Strategy}

As a guide in the identification of plausible feedback strategies and their
limitations, it is useful to restate the problem somewhat in the language of
control systems. To this end, we begin by setting aside the issue of axial
motion and treating the atom as a particle in a cylindrically symmetric,
approximately Gaussian two-dimensional potential whose depth is controlled
by the input light intensity:
\begin{equation}
U\approx -U_{0}e^{-\rho ^{2}/w^{2}}.  \label{2d_effpot}
\end{equation}%
Note that the potential waist $w$ is not simply equated with the previously
introduced cavity field waist $w_{0}$ or with the mode intensity waist $%
w_{0}/\sqrt{2}$, but rather is set by the self-consistent interaction of
atom and light field in the strong coupling regime. Whereas the cavity mode
profile is exactly Gaussian, $U$ is only approximately Gaussian and has an
exact form that is nonanalytical as determined by steady-state solutions to
the master equation for an atom at each value of $\rho $. The potential
depth $U_{0}$ depends on the intensity of the optical field used to drive
the cavity mode. The potential waist $w$ is in fact a (slowly varying)
function of the drive strength as well \cite{acdpra}.

The trapped atom is also subject to friction and to dynamical
noise (momentum diffusion), both arising from decays and
re-excitations of the system on timescales faster than the motion.
In the regime of Ref. \cite{acm}, the contribution of friction is
small compared to the momentum diffusion terms in the equation of
motion.

Because the two-dimensional potential is symmetric, the atom's angular
momentum $L$ is constant, or rather, varies only due to dynamical noise. We
can thus write a one-dimensional effective potential in the $\rho$
dimension,
\begin{equation}  \label{1d_effpot}
V_{eff} = -U_{0}e^{-\rho^2/w^2} + L^2/2m\rho^2
\end{equation}
and thus an equation of motion (for an atom of mass $m$)
\begin{equation}  \label{1d_eqmot}
\ddot{\rho} = -\frac{2\rho U_{0}}{mw^2}e^{-\rho^2/w^2} + \frac{L^2}{m^2\rho^3%
}
\end{equation}
which we notationally simplify to the form
\begin{equation}  \label{1d_eqmot_simp}
\ddot{\tilde{\rho}} = -\tilde{\rho} Ee^{-\tilde{\rho}^2} + \frac{\tilde{L}}{%
\tilde{\rho}^3}
\end{equation}
where $\tilde{\rho}$ is dimensionless ($\tilde{\rho}=\rho/w$), $%
E=2U_{0}/mw^2 $ is the input we control by varying the driving
field strength, and $\tilde{L}=L^2/m^2w^4$ is constant except for
the influence of friction and dynamical noise.

The measurement of light transmitted through the cavity, $T(t)$, is
equivalent to a (noisy) measurement of $\rho (t)$. The noise of this
measurement is largely fundamental quantum noise (shot noise) of detection.
The mapping between $T$ and $\rho $, derived again from steady-state
solutions of the master equation for the coupled atom-cavity system, is not
linear and furthermore depends on the value of the driving strength $E$. The
initial objective is to circularize the two-dimensional orbit -- in other
words, to make $\rho $ constant or to hold $\dot{\rho}=0$ by varying the
control input $E$.

\begin{figure}[htbp]
\includegraphics[width=8.6cm]{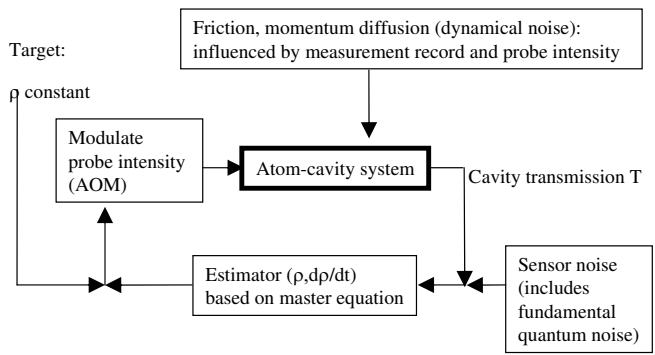}
\caption{Block diagram for the atomic position feedback loop, illustrating
sources of noise and system nonlinearities.}
\label{fig:boxdiagram}
\end{figure}

The simplified system can be described by a block diagram as shown in Figure %
\ref{fig:boxdiagram}. The system exhibits myriad nonlinearities; for
example, ($T\rightarrow \rho $) is nonlinear and depends on $E$, the
dynamical noise depends on $E$, and the equation of motion for $\rho $ is
itself nonlinear. Nonetheless, while this statement of the problem does not
suggest provably optimal feedback strategies, it does motivate some
conceptually simple algorithms based on switching between discrete values of
driving strength $E$. Switching strategies of this sort are often invoked
for the sake of robustness, a major consideration in this scenario;
robustness to dynamical and measurement noise is certainly important, but
perhaps even more relevant is robustness to small uncertainties in system
parameters (e.g., driving strengths or detunings). Switching or
\textquotedblleft bang-bang" type algorithms \cite{controltheory} have the
additional virtue of admitting easy implementation in simulations and in
experimental design.

\vspace{0.7in}

\begin{figure}[tbph]
\includegraphics[width=8.6cm]{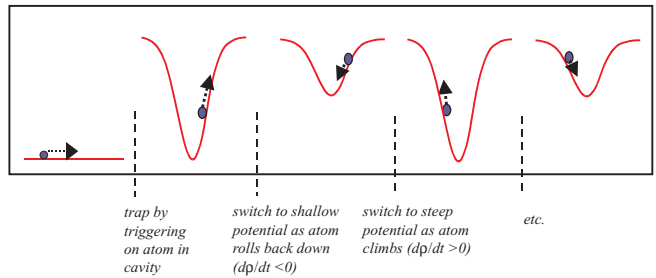}
\caption{General feedback strategy for atomic radial coordinate.}
\label{fig:generalstrategy}
\end{figure}

The feedback algorithms we investigate in this paper all share the same
basic strategy of switching the driving field intensity between two discrete
levels. This corresponds to switching between two potential depths (and,
incidentally, two different sets of friction and momentum diffusion
coefficients as well). The simple objective is to time this switching
relative to the atomic motion so that an atom sees a steep potential when
climbing out of the trap ($\dot{\rho}>0$) and a shallow potential when
falling back towards the trap center ($\dot{\rho}<0$), as illustrated in
Figure \ref{fig:generalstrategy}. The feedback algorithm, then, is based on
switching the potential at turning points of $\rho$, i.e., each time $\dot{%
\rho}$ crosses zero. Implemented effectively, this approach promises
significant dynamical cooling of the radial motion ($\rho\rightarrow
constant $ or $\dot{\rho}\rightarrow0$) in just a few oscillations.

The initial detection and trapping of an atom are accomplished as in Ref.
\cite{acm,acdpra}; a weak probe at driving level \textit{exlo} is used to
detect the atom's arrival in the cavity with minimal effect on the motion,
and an increase in transmission of this beam triggers a switch to driving
level \textit{hi} to populate strong-field-seeking states and trap the atom.
Feedback is then implemented by switching the trapping potential between the
\textit{hi} level and an intermediate \textit{lo} setting, with switching
times based on real-time information about the motion of the single atom.

The simplest algorithm would be to switch back and forth between \textit{hi}
and \textit{lo} potentials at the turning points of $\rho(t)$, which are the
zero-crossings of $\dot{\rho}(t)$. That is, trap initially in \textit{hi},
switch to \textit{lo} when $\dot{\rho}$ crosses zero from above (i.e., when $%
\rho$ begins to decrease), switch back to \textit{hi} when $\dot{\rho}$
crosses zero from below ($\rho$ reaching its minimum and increasing), and so
on until the atom escapes. However, this strategy calls for a theoretically
infinite sequence of switching events, while it is desirable to instead
achieve a steady state in some long-time limit. The presence of dynamical
noise implies that the exact steady state of $\dot{\rho}\rightarrow0$ is in
any case unreachable, so we replace it with a goal of confining $\dot{\rho}$
to some range $[-lim,+lim]$. Thus the feedback strategy is modified to
include slight hysteresis: \textit{lo}$\rightarrow$\textit{hi} when $\dot{%
\rho}\rightarrow +lim$ from below, \textit{hi}$\rightarrow$\textit{lo} when $%
\dot{\rho}\rightarrow -lim$ from above. With this modification, switching
stops once $\dot{\rho}$ is confined within the acceptable range.
Furthermore, we prefer a steady state with \textit{hi} potential for reasons
of the deeper confinement; to bias the system towards this final state, we
use asymmetric hysteresis limits: \textit{lo}$\rightarrow$\textit{hi} when $%
\dot{\rho}\rightarrow +lim$ from below, \textit{hi}$\rightarrow$\textit{lo}
when $\dot{\rho}\rightarrow -(lim+\delta)$ from above.

\section{Simulations of Feedback Algorithms in Operation}

For our simulations, we first choose driving strengths
$n_{exlo}=0.05$ photons in the empty cavity, $n_{hi}=0.3$ photons
in the empty cavity, and $n_{lo}=0.15$ empty-cavity photons. These
driving strengths are high enough so that an atom of typical
kinetic energy can be trapped by both \textit{lo} and \textit{hi}
drives, yet low enough to ensure the increase in momentum
diffusion between \textit{lo} and \textit{hi} does not outstrip
the increase in potential depth.

Our simulations of the atom-cavity dynamics are based on the
treatment described in detail in Ref. \cite{acdpra,acd}; the
treatment is fully quantized for the atomic internal state and the
cavity light field, but considers the atomic center-of-mass motion
quasiclassically. This approximation is suitable for the current
experimental situation, with more manifestly quantized motion to
be accessed by better cooling and/or detection of the atom's axial
motion. The non-conservative terms in the system, in the form of
friction and momentum diffusion, are included in the simulation;
the resulting ``heterodyne transmission" trace is a perfect record of $%
|\langle a \rangle|^2$, on which measurement bandwidths and shot noise must
be imposed separately. Shot noise is modelled as Gaussian white noise with
an amplitude that depends on the size of the (noiseless) transmission signal.

In the presence of sensor noise, we require estimators for the quantities $%
\rho$ and $\dot{\rho}$. Because one parameter ($L$) in the equation of
motion is unknown and in fact slowly varying, we have chosen not to
implement estimators based on a Kalman-filter approach. More sophisticated
treatments include Kalman-type approaches to simultaneously estimate $\rho$,
$\dot{\rho}$, and $L$, but these have not been explored in full detail.
Meanwhile, we choose to estimate $\rho(t)$ and $\dot{\rho}(t)$ directly from
the measurement record, with no explicit reference to the equation of motion
for the system.

The noisy transmission signal $T$ is sampled at 1MHz as in the
experiment of Ref. \cite{acm}. To estimate $\rho$, we first
perform an RC low-pass filter on $T$ at 100kHz. This step is an
infinite impulse response (IIR) filter which introduces only a
small delay in the estimator. This filtered transmission signal is
then put through a lookup table with linear interpolation to
obtain $\rho_{est}(t)$.

The resulting $\rho_{est}(t)$ tracks the actual $\rho$ closely but still
with significant noise. Obtaining a time derivative without excessive noise
thus requires some care. A variety of methods are mentioned in Ref. \cite%
{hpreport}, in which the authors are concerned with estimating the
sign of a time derivative in order to feed back to a system --
essentially the same problem we encounter. We employ a simple
finite impulse response (FIR) filter that takes the slope of a
linear least squares fit to $\rho_{est}(t)$ over a window of fixed
size. A detailed implementation of this filter is found in Ref.
\cite{vainio}. The resulting $\dot{\rho}_{est}(t)$ is a good
estimator for $\dot{\rho}$ at the middle of the window, so the
delay induced is approximately half the window size. We find via
numerous simulations that a window size of 30 to 40 $\mu$s gives a
signal $\dot{\rho}_{est}(t)$ which is quiet enough for use in our
control. Thus reliable estimation in the presence of noise
introduces a delay of approximately 15 to 20 $\mu$s in the
feedback loop. This time
delay can be compared to a typical atomic orbital period of $%
\tau_{r}\sim100\mu$s, corresponding to a period of $\sim 50\mu$s for $\rho$.
Feeding back effectively in the presence of such large delays requires a
certain amount of adjustment to the naive cooling algorithm, as discussed in
detail below.

\subsection{Actual Dynamics but No Measurement Noise}

\begin{figure}[htbp]
\includegraphics[width=8.6cm]{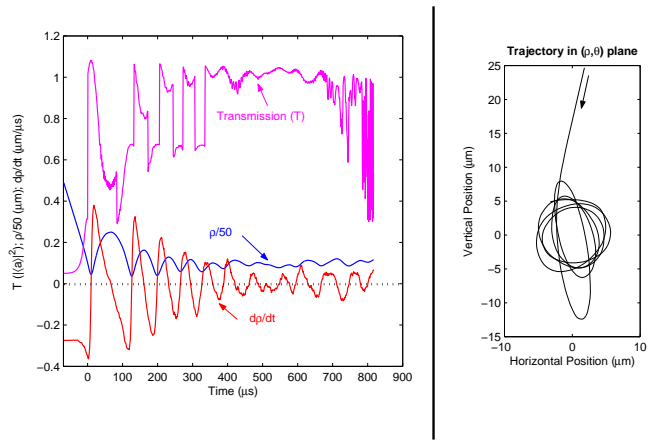}
\caption{Simulated feedback example with perfect sensing of
$\protect\rho(t)$. The graph on the left shows idealized cavity
transmission $T(t)$ (purple),
  radial position $\rho (t)$ (blue), and $d\rho/dt$ (red) for a
  simulated atom trajectory.  In the right panel the same atomic
  trajectory is shown in the $(x,y)$ (or $(\rho,\theta)$) plane.
  This trajectory begins as a nearly vertical transit through the
  cavity, but is circularized by the feedback employed.  In the left
  panel, the circularization is evident as $d\rho/dt$ is damped towards
  zero and $\rho$ and $T$ become nearly constant at the end of the
  trajectory.} \label{fig:egnonoise}
\end{figure}

Before treating the case of actual experimental noise, we explore the
performance of our feedback strategy in simulations with noiseless
measurement and thus perfect, zero-delay sensing of $(\rho,\dot{\rho})$.
Figure \ref{fig:egnonoise} shows an example trajectory using this
asymmetric-hysteresis switching strategy. The values of cavity transmission $%
T$ and atomic position variables are sampled every 1$\mu$s, but the
dynamical timestep is 3,000 times finer than this ``information" timestep.
Note that axial motion (the $x$ direction our strategy neglects) is included
in the simulation, and when its amplitude is large it gives rise to the very
fast variations seen in $T(t)$. However, since the period of $x$ motion is
similar to the information timestep used, note that these signals are
undersampled in the record.

A 10-$\mu$s box filter is applied to $\dot{\rho}$ in order to
remove some oscillations caused by $x$ motion and also partially
to anticipate some effects of noisy detection and delay. In
setting the conditions for potential-switching, we employ the
asymmetric hysteresis described above with
$(lim,\delta)$=(0.05,0.03)$\mu$m$/\mu$s, so the potential depth is
switched at $\dot{\rho}=+0.05,-0.08 \mu$m$/\mu$s. Switching
events, since they correspond to turning the light level up and
down, can be seen as sharp edges in the transmitted light $T$. As
the example illustrates, the control strategy successfully
circularizes atomic trajectories within a few orbital cycles. This
can be seen from $\rho(t)$ as well as from the trajectory shown in
the $(\rho,\theta)$ plane. The hysteresis limits are chosen so
that variations in $\rho$ due only to dynamical noise tend not to
trigger any switching of the drive. This is
illustrated by the continued high control level throughout the time $%
t=500-650\mu$s, while $\rho$ is wandering diffusively rather than
oscillating with regularity.

The overall trap lifetime is dominated, as in this example, by heating in
the $x$ direction. (In the example shown, note the fast, large-amplitude
variation in transmission just before the atom escapes; this is a signature
of rapid axial heating.) Thus our feedback strategy has little impact (at
the level of 10\%) on average trapping lifetimes. Circularizing the orbit
helps decrease axial heating since the potential depth no longer wanders as $%
\rho$ varies; however, the feedback is accomplished by sharp
switching events which occur at arbitrary times relative to the
oscillations in the $x$ direction. The overall impact on lifetimes
is therefore small in the simulations we have performed. Since the
feedback algorithm is aimed at reducing motion in the $\rho$
direction, its success is best measured by its performance at that
task specifically. Lifetime effects can become apparent only if
the axial motion is suppressed by some other means; that case is
treated briefly in Section E below.

\subsection{Adding Measurement Noise Adds Delays}

\begin{figure}[htbp]
\includegraphics[width=8.6cm]{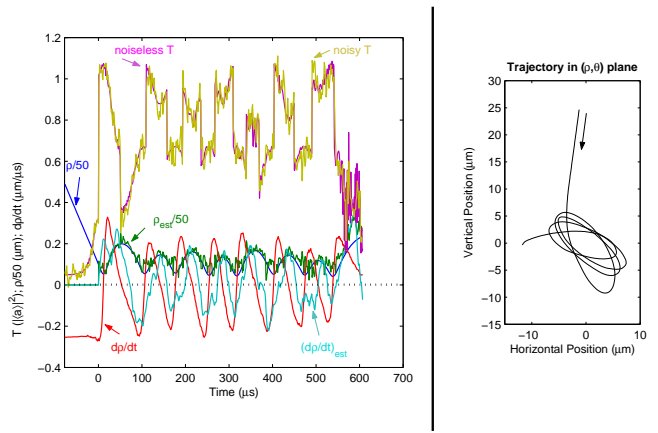}
\caption{Measurement noise leads to filtering and unacceptable loop
delay in this example.  In addition to the quantities plotted in
Figure \ref{fig:egnonoise}, the left panel now displays the
experimentally realistic transmission $T(t)$ \textit{with} noise
(yellow), and the radial motion estimators $\rho_{est} (t)$ (green)
and $(d\rho/dt)_{est}$ (light blue) arising from the noisy
transmission signal.  Feedback switching is triggered by the value
of $(d\rho/dt)_{est}$, but since the estimator lags well behind the
true $d\rho/dt$ damping is not achieved, and indeed the
two-dimensional trajectory shows little change in shape over time.}
\label{fig:egbaddelay}
\end{figure}

The addition of measurement noise and consequent estimation of $\dot{\rho}$
introduce significant loop delays, as described above. Since $\dot{\rho}%
_{est}$ can be almost half a cycle behind the actual $\dot{\rho}$, we expect
naive switching to be well out of phase with the atomic motion and thus
relatively ineffective as a cooling mechanism. Figure \ref{fig:egbaddelay}
shows an example in which the feedback strategy is identical to that of
Figure \ref{fig:egnonoise}, but applied to $\dot{\rho}_{est}$ rather than to
$\dot{\rho}$ itself. The resulting time delay seriously compromises
performance, as shown. In the figure, note that the switching events,
recognizable as sharp edges in transmission, do not line up with turning
points of $\rho$. As a result, $\dot{\rho}$ is not damped and the trajectory
remains elliptical.

\subsection{Account for Delays by Waiting a Cycle}

Since it seems clear we cannot simply close the loop with the delays
necessitated by measurement noise, we choose to address the problem by
adding even more delay -- that is, by detecting a switching condition ($\dot{%
\rho}_{est}$ crossing a hysteresis limit) and then waiting to switch the
potential at a time which should catch the \textit{next} oscillation of $%
\rho $. A first attempt in this direction would be to assume a fixed period
for oscillations of $\rho$. In this case the additional wait before
switching is given by this fixed period minus the estimator delay for $\dot{%
\rho}_{est}$. Since each switching time is now set by the detected signal
from the previous cycle, the first switching opportunity (first minimum of $%
T $ and maximum of $\rho$) will be missed in this strategy. Rather than miss
this cooling cycle, we impose a single switching event a fixed time after
the initial trap turn-on. Thus the potential switches \textit{exlo}$%
\rightarrow$\textit{hi} on the initial trigger, \textit{hi}$\rightarrow$%
\textit{lo} a fixed time later, and \textit{lo}$\leftrightarrow$\textit{hi}
thereafter based on the last zero-crossing time of $\dot{\rho}_{est}$.

\begin{figure}[htbp]
\includegraphics[width=8.6cm]{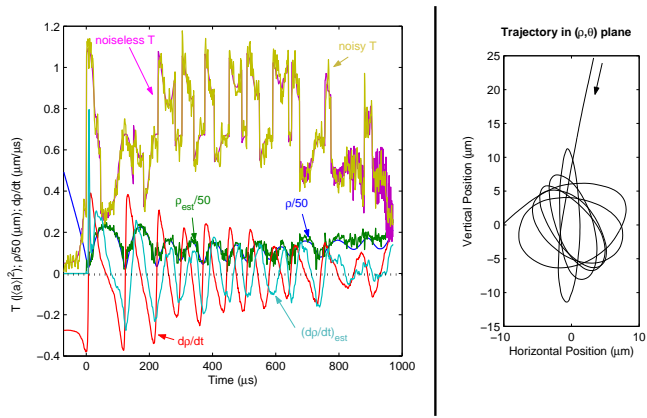}
\caption{In this example delay is dealt with by tracking
$\protect\rho$ turning points and using this information from each
cycle to switch the potential at the predicted \textit{next} turning
point.  This strategy give moderate success at damping $d\rho/dt$
and circularizing the trajectory.} \label{fig:egscweather}
\end{figure}

However, the actual dynamical period varies by easily a factor of
two over the course of an atom's trapping lifetime due to changing
amplitude of oscillation in the anharmonic potential, as seen in
Figure 3b. Thus the fixed-period assumption is a poor one. A
better strategy is to record the length of each period in
$\dot{\rho}_{est}$ and assume each cycle will be the same length
as the previous recorded one. Thus the \textquotedblleft waiting
time" estimate will adjust itself as the dynamical period changes,
though it will in general be one cycle behind. This strategy is
employed for the trajectory shown in Figure \ref{fig:egscweather}.
The initial switch occurs 45$\mu $s after trap turn-on, the
least-squares window
is 40$\mu $s, and the \textquotedblleft wait time" between subsequent $\dot{%
\rho}_{est}$ limit-crossings and the resultant potential switches is given
by the previous period minus 20$\mu $s. This switching strategy, with
deliberate delay based on an active measurement of the $\rho $ oscillation
time, appears to be a viable means of performing control in the presence of
sensor noise and its associated loop delay.

\subsection{Comparisons with Open Loop Strategies}

To evaluate the effects of feedback more quantitatively, we introduce a
figure of merit for the damping of radial oscillations in an atomic
trajectory. Since the goal of the control strategy is to confine $\dot{\rho}$
near zero, the performance can be measured by comparing the variance of $%
\dot{\rho}$ over intervals of equal duration near the beginning of
the trajectory and after feedback has been operating for some
time. We choose a time window of duration 200$\mu$s as long enough
to encompass well over one cycle of the radial motion. The
comparison is taken between two such windows separated from one
another by 200$\mu$s; this delay is selected as long enough for
several cycles of feedback action, yet short enough so that the
statistic exists for a large fraction of trapped atom events.
Finally, we base our performance measure on the experimentally
accessible $\dot{\rho}_{est}$ rather than on $\dot{\rho}$ itself.
Thus our figure of merit for feedback performance is given by
\begin{equation}  \label{figmerit}
M=\frac{\sigma^2_{15\mu s\rightarrow 215\mu s}(\dot{\rho}_{est}(t))}{%
\sigma^2_{415\mu s\rightarrow 615\mu s}(\dot{\rho}_{est}(t))}
\end{equation}
where times are measured from the initial trapping (\textit{exlo}$%
\rightarrow $\textit{hi}) switch. Large values of the quantity $M$
correspond to well-damped radial motion, $\rho(t) \rightarrow
constant$, though orbits may still be circular at any radius $\rho
\geq 0$. (Damping in the sense of actual energy removal is
discussed explicitly in Section E.) Small ($\sim 50-100\mu s$)
changes in delay time or window size have been investigated and do
not appreciably change the nature of the results for $M$.

\begin{figure}[htbp]
\includegraphics[width=8.6cm]{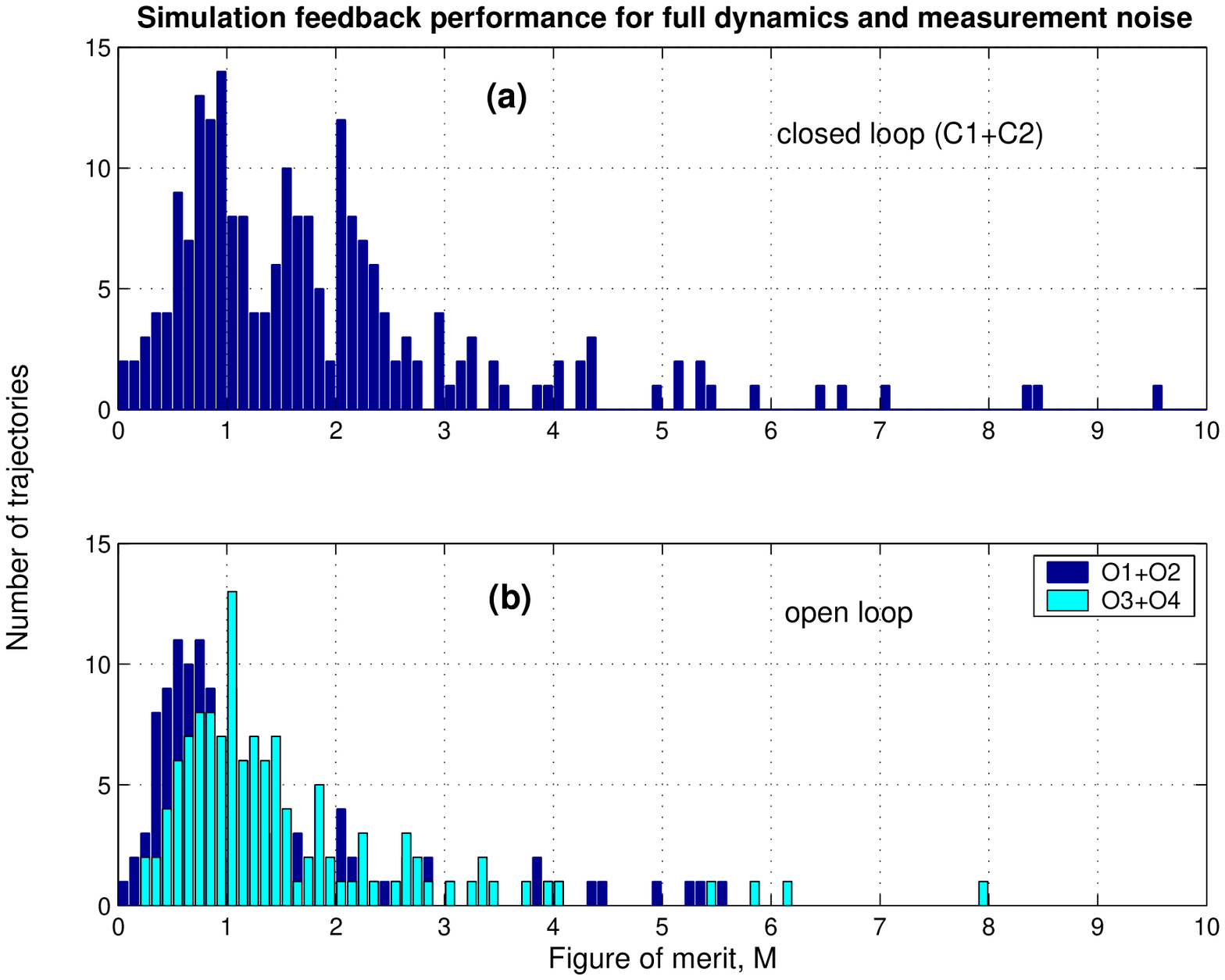} \vspace{0.2in} %
\includegraphics[width=8.6cm]{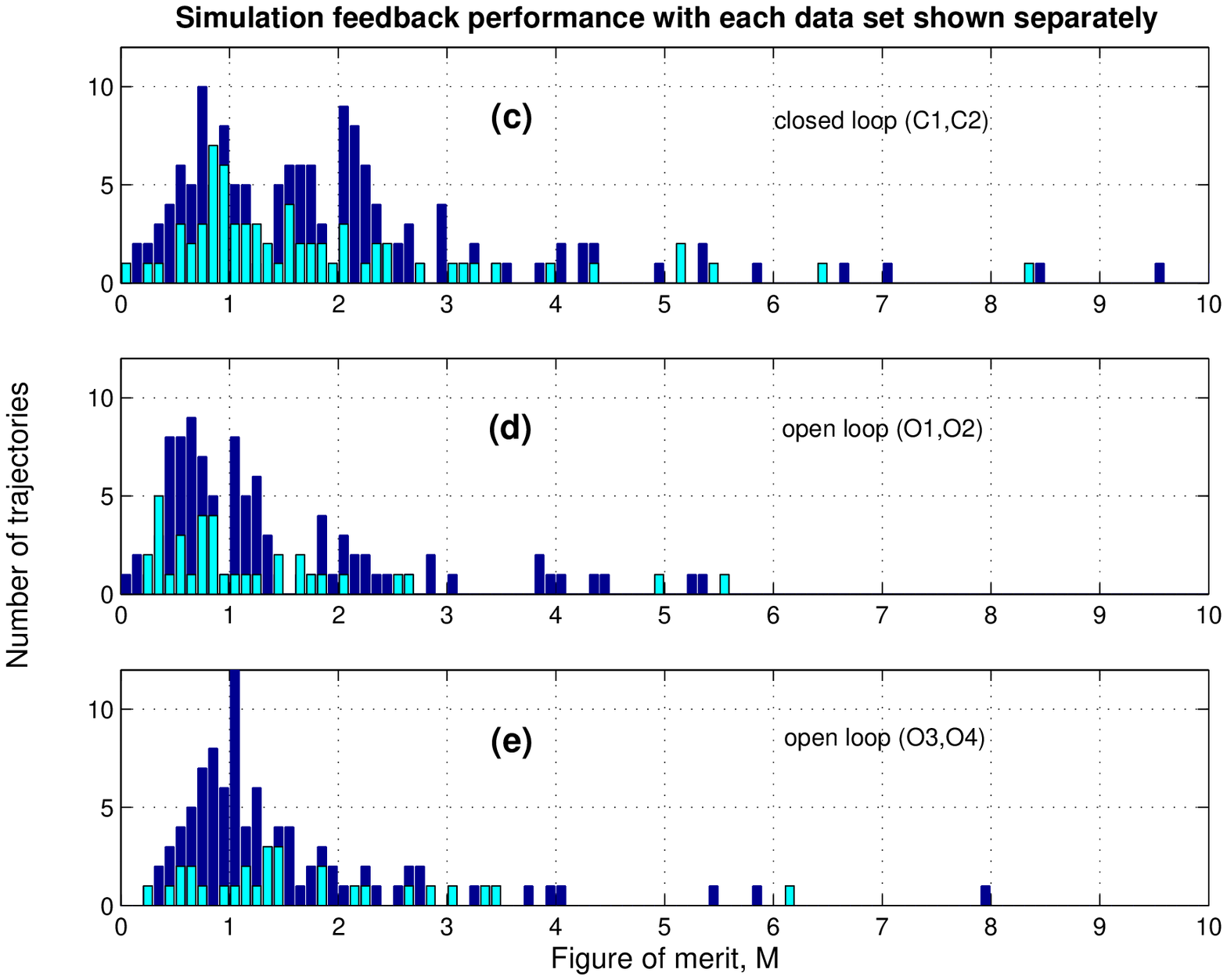}
\caption{Figure of merit for feedback in closed- and open-loop cases
(full dynamics). $M>$1 indicates damping of radial motion.}
\label{fig:merithistograms}
\end{figure}

Figure \ref{fig:merithistograms} shows histograms of $M$ for several data
sets in which different switching protocols, detailed in Table \ref%
{table:datasets}, have been employed. Each data set is generated by
simulating a fixed number of individual atom drops from the known
distribution of initial conditions. Only some fraction of trajectories
result in a triggering/trapping event, and of these only a fraction of atoms
have dwell times long enough to compute a value for $M$. Thus, for example,
set C2 was generated from 5,000 trajectories, yielding 1,335 trigger events
and 147 trajectories for which $M$ could be obtained (i.e., with dwell times
at least 615$\mu$s).

\begin{table}[h]
\centering
\begin{tabular}{|llll|}
\hline
Data set & switching protocol & \# atom drops & \# trapped \\ \hline\hline
C1 & Closed loop & 2000 & 534 \\
& Switch once &  &  \\
& 45 $\mu$s after &  &  \\
& initial trigger. &  &  \\
& Thereafter switch &  &  \\
& (previous cycle length) &  &  \\
& - 20$\mu$s after &  &  \\
& $(d\rho/dt)_{est}$ &  &  \\
& crosses limits. &  &  \\ \hline
C2 & Same as above (C1) & 5000 & 1335 \\ \hline
O1 & Open loop & 2000 & 561 \\
& Switch every 45$\mu$s &  &  \\
& after initial trigger. &  &  \\ \hline
O2 & Same as above (O1) & 5000 & 1319 \\ \hline
O3 & Open loop & 2000 & 552 \\
& Switch every 35$\mu$s &  &  \\
& after initial trigger. &  &  \\ \hline
O4 & Same as above (O3) & 5000 & 1322 \\ \hline
\end{tabular}%
\caption{Exact conditions used for data of the full-simulation histograms.}
\label{table:datasets}
\end{table}

While Table \ref{table:datasets} gives the specifics of each data
set represented in Figure \ref{fig:merithistograms}, the essential
comparison is between closed loop -- i.e., active feedback --
algorithms and open loop counterparts which simply switch potentials
in a predetermined sequence independent of real-time position
information for the individual atom. The closed loop algorithm is
that of Figure \ref{fig:egscweather}, in which measurement noise and
loop delays are dealt with by waiting nearly one cycle to apply the
knowledge of motion gained during the previous oscillation. The open
loop algorithms, in contrast, simply switch the potential between
\textit{hi} and \textit{lo} at fixed intervals following the initial
trapping event; the fixed interval is chosen to coincide with a
reasonable average value for an atomic oscillation period.  Over a
long atomic trajectory, the atomic trajectory clearly evolves out of
phase with any one open loop switching algorithm; since this
evolution is different for each atomic trajectory, we compare
closed-loop with open-loop strategies as a means of evaluating the
importance of \textit{real-time} measurement and feedback in our
algorithm.

Closed loop, active feedback clearly damps radial oscillations more
effectively than its open loop counterpart. The $M$ distributions in Figure %
\ref{fig:merithistograms}(a,c) have larger mean than those in Figure \ref%
{fig:merithistograms}(b,d,e), with the closed-loop histograms showing many
trajectories pushed out to higher values of $M$ by the active feedback.
These results indicate that real-time measurements of $\rho(t)$ can indeed
be applied to facilitate cooling of a single atom's motion in that dimension.

Further refinements should improve the performance of the algorithm. For
instance, the cycle-length predictor could be changed to allow asymmetries
between $\dot{\rho}>0$ and $\dot{\rho}<0$ half-cycles. Additionally, the
filters themselves could be adjusted or replaced with better estimators
which incorporate information about angular momentum $L$.

The simulations reported here could in principle be employed, with
minor modifications, to address the experimental regime of the
atom-cavity feedback performed in Ref. \cite{rempefb}.  We have not
attempted a quantitative comparison since results may be sensitive
to details of the experimental implementation. However, we note that
because of the values of $(g, \kappa)$ for the optical cavity
employed, Ref. \cite{rempefb} was carried out in a qualitatively
different regime from the one in which our simulations operate.  In
particular, while both scenarios involve a measurement bandwidth
which averages over axial motion, in the case of Ref. \cite{rempefb}
the amplitude of the axial motion is in fact quite large. One key
observation in that experimental setting was in fact the change in
measured transmission as atoms either oscillated within a single
standing-wave antinode or ``flew" across multiple antinodes. Without
the measurement bandwidth to observe axial oscillations directly,
one is hard pressed to separate axial modulation from radial motion
in the manner presented here.  While both feedback algorithms
involve a similar switching of cavity driving strength, the feedback
modeled here is directed at cooling a \textit{particular component}
of the atom's motion $\vec{r}(t)$. Figures \ref{fig:merithistograms}
and \ref{fig:noax} illustrate the damping of $d\rho/dt$.  By
contrast, the lifetime enhancement of Ref. \cite{rempefb} occurs
because the authors are able to discern when the atom is trapped at
high $g(\vec{r})$ and selectively turn down the otherwise large
diffusive \textit{heating} by lowering the trap intensity at those
times.

\subsection{Performance with Axial Motion Suppressed}

In a final set of simulations, we investigate the performance of
our radial cooling algorithm in a setting where the axial atomic
motion is independently suppressed. With no (or little) axial
motion, axial heating no longer limits trapping times and the
effects of radial feedback can be seen more clearly. To achieve
this in simulations, we impose an \textit{ad hoc} elimination of
the axial dimension; however, this case could be relevant to
several future experimental scenarios. For example, trapping and
sensing mechanisms, both currently accomplished with the same
probe beam, could be separated to allow a trapping field with a
low scattering rate and much-reduced axial diffusion.
Alternatively, the separation of axial and
radial timescales could be exploited, either to apply axial cooling\cite%
{stevencooling} between cycles of radial feedback or to simply avoid extra
axial heating by ramping the potentials up and down at a rate that appears
adiabatic to the axial motion while still impulsive in the radial dimension.

Prospects for implementation aside, simulations with no axial motion
demonstrate some aspects of the radial feedback protocol that are otherwise
less transparent. We explore this regime with a set of simulations that
differ in three ways from those presented above. First, the axial dimension
is eliminated entirely and the atom is artificially constrained to remain at
rest at an antinode of the standing-wave cavity field. Second, since axial
heating is no longer an issue, we employ somewhat deeper trapping potentials
than in the simulations above. The weak probe level is still $n_{exlo}=0.05$
photons in the empty cavity, but now we turn on the trap initially at a
level $n_{exhi}=0.6$ photons in the empty cavity, and we feed back by
switching between this and the weaker level $n_{hi}=0.3$ photons. The
effective potentials thus generated are $\sim 50\%$ deeper than in the
previous simulations using $n_{hi}$ and $n_{lo}$. Finally, without axial
motion we employ a coarser computational timestep of (1/30)$\mu$s.

\begin{figure}[htbp]
\centerline{(a)}\centerline{\includegraphics[width=8.6cm]{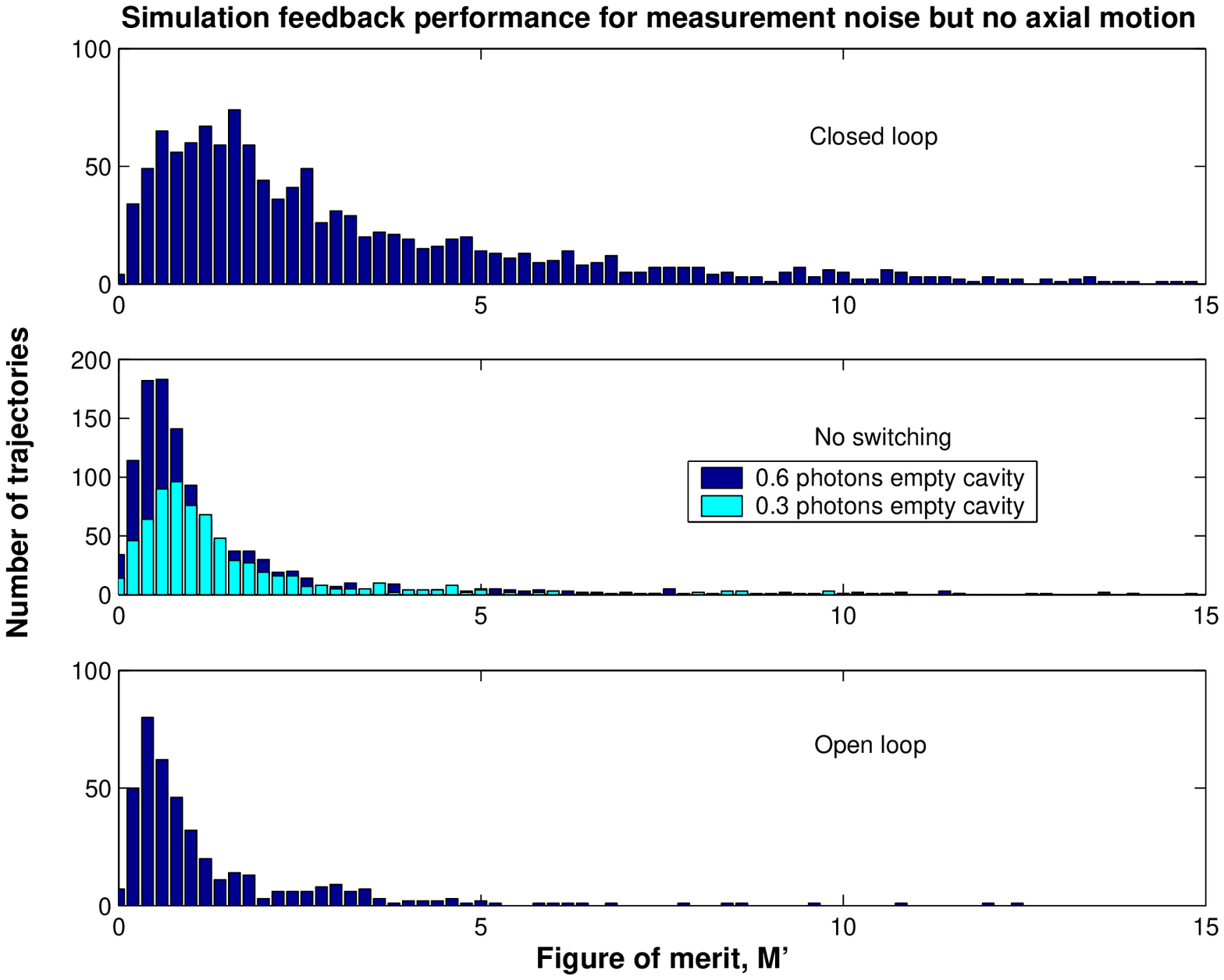}}
\vspace{0.2in} \centerline{(b)}\centerline{%
\includegraphics[width=8.6cm]{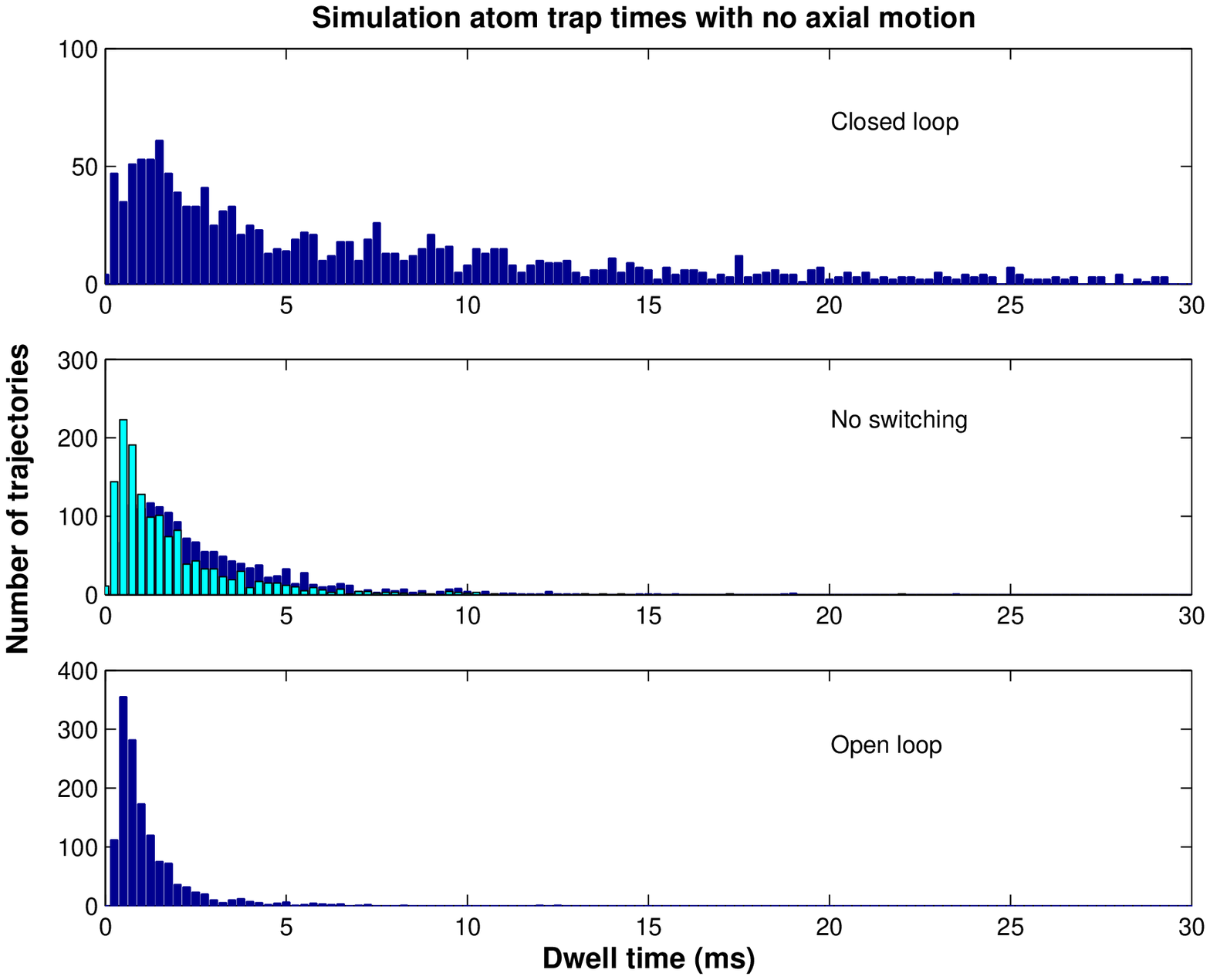}}
\caption{Feedback performance and lifetime enhancement in
simulations with no axial motion.  $M^{\prime}>$1 indicates damping
of radial motion.} \label{fig:noax}
\end{figure}

Figure \ref{fig:noax}(a) shows the feedback figure of merit for the cases of
closed-loop feedback, constant trapping at $n_{hi}$ or $n_{exhi}$, and
open-loop switching. The open-loop protocol in this case is to trap
initially with $n_{exhi}$ and switch between driving levels $n_{exhi}$ and $%
n_{hi}$ every 40$\mu$s during the transit duration. Taking advantage of
longer overall atom dwell times, we can now consider time windows separated
by a greater delay, so the quantity displayed here is
\begin{equation}
M^{\prime }=\frac{\sigma _{15\mu s\rightarrow 215\mu s}^{2}(\dot{\rho}_{est}(t))}{%
\sigma _{1015\mu s\rightarrow 1215\mu s}^{2}(\dot{\rho}_{est}(t))}
\label{eq:figmerit2}
\end{equation}%
rather than the original $M$ of Figure \ref{fig:merithistograms}. The result
is qualitatively similar to that of the full simulation, with the closed
loop strategy performing significantly better than its open loop
counterpart. In this case, the mean value of $M^{\prime }$ for closed-loop
feedback is $\sim 2.5$ times greater than for open-loop switching or for no
switching.

The data of Figures \ref{fig:merithistograms} and
\ref{fig:noax}(a) indicate that our feedback algorithm acts to
drive $\dot{\rho}$ to zero, i.e., to circularize atomic orbits at
a constant value of $\rho $.  If angular momentum $L$ is not
correspondingly increased, this effect implies a damping of radial
energy. However, we may investigate more directly whether this
algorithm actually removes total energy from the radial motion.
With axial motion eliminated, we can now explore this issue by
asking how the feedback algorithm affects trapping lifetimes.
Figure \ref{fig:noax}(b) shows atom dwell times for the same three
cases of closed-loop feedback, constant drive levels of $n_{exhi}$
or $n_{hi}$, and open-loop switching. The increase in lifetime for
closed-loop feedback is immediately apparent. Indeed, the
closed-loop results agree well with an exponential lifetime of
8.9ms, as contrasted with 2.6ms for trapping at $n_{exhi}$ alone,
1.9ms for trapping at $n_{hi}$, and 1.1ms for open-loop switching
between the drive levels.

In the closed-loop case the trap potential is varied during the
transit but never exceeds the depth associated with driving level
$n_{exhi}$. Nevertheless, trap times exceed those for constant
driving at $n_{exhi}$, demonstrating that active feedback as applied
to the radial dimension does act to remove radial energy, in
addition to simply pinning $\rho $ to a constant value. The same
point can be illustrated by considering the change in total energy
between the beginning (15-215 $\mu$s after trigger) and end
(1015-1215 $\mu$s after trigger) of an atom's dwell time in the
cavity; the active feedback strategy produces a modest ($\simeq$10\%)
net energy removal not seen under either the simple trap or the
open-loop switching protocol.  Since the feedback algorithm performs
better than both open-loop and fixed-trap strategies when measured
by dwell time, radial damping $M^{\prime}$, \textit{or} total energy
removal, we characterize its effect as in fact actively cooling a
component ($\rho(t)$) of the atomic motion.

Note that all lifetime in the two-dimensional simulation are
enhanced relative to the full three-dimensional case, in which both
experiments and simulations have mean trapping times of only $\sim
400\mu $s. Lack of lifetime enhancement from radial feedback in the
full simulation is some indication of the very weak mixing between
axial and radial motion, so that cooling in one of these dimensions
does little to control temperatures in the other.

\section{Outlook for Experimental Implementation}

The feedback simulations discussed in this paper have been conducted with
very close reference to the conditions realized in the experiment of Ref.
\cite{acm}, in particular for cavity properties, trapping statistics, and
signal-to-noise in the balanced heterodyne detection. The current
experimental effort, aimed at realizing the feedback strategies described
here, employs very similar conditions while incorporating some improvements
as in Ref. \cite{fort,mckeever,mcknature}. Notable changes from Ref. \cite%
{acm} include a slightly shorter cavity, improved vacuum pressure in the
cavity region ($3$x$10^{-10}$ torr) enabled by a differentially pumped
double chamber, and cavity length stabilization via an error signal
generated by an independent laser one free spectral range away from the
cavity QED light. Implementation of the feedback strategy described above is
clearly outside the regime of analog electronics, so an additional
modification is the use of digital processing and FPGA technology\cite{fpga}%
. With these tools, experimental data similar to the simulation results
presented above seem well within reach.

It seems reasonable to ask how much experimental data should be necessary to
exhibit a distinction between active feedback and open loop schemes. From
the simulations of Figure \ref{fig:merithistograms} (Table \ref%
{table:datasets}), we see that with data sets of about 500 trapped atoms the
differences between open- and closed-loop schemes already begin to become
apparent, and these differences are well demarcated with two or three times
that much data. With fairly conservative estimates of one trigger per MOT
drop and one MOT drop every 5 seconds, this means significant effects could
well be seen with just one to two hours of experimental data at each
setting. Much more data collection is experimentally realistic, allowing
exploration of a wealth of additional questions.

With the atom-cavity system's capacity to give real-time information
on a trapped atom's position, active feedback might seem to be an
attractive enabling technology for experiments that require a
stationary atom at fixed $g(\vec{r})$ in the cavity field.  The
present work was undertaken in part to explore this option through
simulation of realistic experiments.  While we conclude that
feedback of a uniquely real-time nature can measurably alter
properties of the atomic motion, we also find that measurement
bandwidths and signal to noise strongly constrain the precision of
feedback performance under current experimental conditions. Feedback
experiments address important topics in quantum measurement and
control; meanwhile, trapping strategies using auxiliary fields offer
a more direct route towards a stationary atom in a cavity.

\section{Current Limits and Future Directions}

The feedback algorithm developed above for the atomic radial position $\rho$
is subject to basic limits arising from dynamical and measurement noise in
our system. These limits can be expressed as lower bounds on the
(one-dimensional) temperature for $\rho$. Because the strategy is based on
discrete switching, with feedback delayed and timed based on the previous
switching-cycle length, the control is always based on information gathered
over the previous motional cycle. Thus dynamical noise over an atomic
motional timescale will set a lower limit on $T_{\rho}$. Referring to \cite%
{acdpra}, we find that momentum diffusion (due to spontaneous emission)
gives a typical energy increase per radial oscillation time of $\Delta
E_{\tau_{r}} \sim 0.02(U_{0}) \sim k_{B}(50\mu K)$. Furthermore, measurement
noise places a limit on the detectable amplitude of $\rho$ variations. This
amplitude depends strongly on the absolute value of $\rho$ due to the
nonlinearity of the $T \rightarrow \rho$ mapping. However, using the
measured sensitivity from the atom-cavity microscope, we estimate that over
a motional cycle we can resolve $\rho$ oscillations of amplitude $(20nm/%
\sqrt{Hz}) \sqrt{1/2\pi\tau_{r}} \approx 0.77 \mu m$. On the side of the
cavity mode, where the effective potential is steepest, this corresponds to $%
T_{\rho} \approx 150 \mu K$. While this limit corresponds closely to the
simulations of the previous section, where axial motion is suppressed, the
full simulation never reaches this limit because axial heating cuts off
atomic lifetimes too quickly. Thus improvements to address axial heating are
of great interest for seeing the full effect even of radial cooling.

Beyond the experimental and algorithmic variations already discussed, a
number of broader questions are raised by the use of active feedback to
dynamically cool a component of motion for a single atom. One question deals
with the ultimate limits of such a cooling mechanism. Within the current
experimental setting, limits to radial cooling arise from atomic lifetimes
(dominated by axial motion), but are also constrained by the dynamical noise
and by the shot noise of detection. Some lifetime and dynamical noise issues
could be addressed by separating trapping and sensing, for example by using
a far off resonance trap in conjunction with a sub-photon level cavity QED
probe\cite{mckeever}. The remaining issues would then center on
signal-to-noise for the atomic position measurement, as well as on limits
imposed by backaction of the measurement itself as it approaches the
standard quantum limit\cite{sql,gambetta}. These limits must be considered
not only in the context of near-resonant probing, as treated here, but also
in the case of a far detuned probe for which atomic position information is
extracted from probe phase shifts, as first measured in \cite{fullmeas}.

Extension of active feedback beyond the $\rho $ dimension raises related
questions. The question here is one of using various techniques -- for
example, a symmetry-breaking potential as could be provided by a
higher-order transverse mode of the cavity or frequency-domain filtering of
the signal -- to estimate and control a three-dimensional vector using the
time record of a single quantity, the transmitted light field. Undoubtedly,
different driving parameters, detection methods, and data processing will be
appropriate depending on the relative importance placed on information about
each dimension of the motion. These questions address in various ways some
basic issues of optimal state estimation and control for single quantum
systems, and this experimental system promises to be a rich one for
exploring such issues in further detail.

\begin{acknowledgments}
We gratefully acknowledge the contributions of many colleagues, including
Joseph Buck, Dong Eui Chang, Domitilla del Vecchio, Andrew Doherty, Martha
Gallivan, Christina Hood, Ron Legere, Jason McKeever, Dominic Schrader, and
Jun Ye. This work has been funded by the National Science Foundation.
\end{acknowledgments}

\bibliographystyle{plain}
\bibliography{algorithmbib}

\end{document}